# Flyback-Based Multiple Output dc-dc Converter with Independent Voltage Regulation

Mohammad Tahan
*Electrical and Computer Engineering*
*University of Massachussetts Lowell*
Lowell, USA
Mohammad_Tahan@student.uml.edu

David Bamgboje
*Electrical and Computer Engineering*
*University of Massachussetts Lowell*
Lowell, USA
Olorunfemi_Bamgboje@student.uml.edu

Tingshu Hu
*Electrical and Computer Engineering*
*University of Massachussetts Lowell*
Lowell, USA
Tingshu_Hu@uml.edu

*Abstract*—This paper proposes a new single input multiple output power supply by integrating a flyback converter and several buck converters. The flyback converter works as the main regulator, and the buck converters provide series voltage compensation with the aim of tight regulation. A time multiplexing switching scheme is proposed to deliver multiple output voltage levels via a two winding transformer and to eliminate the cross regulation between output channels. This configuration reduces the size of flyback transformer and filter capacitors, and consequently improves the overall form factor. A detailed steady state analysis is conducted on the circuit to obtain the design criteria. A three output channel power supply is designed and the effectiveness of the proposed configuration is validated via simulation with a MATLAB/Simscape model. Simulation results also demonstrate satisfactory transient response to load changes.

*Keywords—Flyback Converter, SIMO Topology, Power Supply, Series Compensation*

## I. Introduction

Application of multiple output dc-dc converters in many fields such as military, biomedical, energy harvesting, renewable energies and electric vehicles (EVs), are in high demand. EV is a good example which employs different loads with different voltage levels including electrical motor, lighting, air conditioning etc. A simple, compact and efficient converter that can satisfy different load conditions is the basic requirement in designing multiple output dc-dc converters. Conventionally, *N* independent single output converters have been adopted to supply *N* different loads. Although the scheme is simple and effective, it usually comes with higher form factor and cost, especially when galvanic isolation is required. The purpose for creating isolation is to ensure safety and to avoid operation interruption from fault conditions, and to enable different ground potentials for two or more electric circuits. Boost, buck, buck-boost and cuk converters all yield non-isolated converter topologies. On the other hand, half-bridge, full-bridge, push-pull, forward, and flyback converters yield isolated converter topologies. Since every isolated converter employs switching mode power supply (SMPS) transformer which is bulky and expensive, using *N* separate single output isolated converters would further increase the cost and size. A flyback converter is a feasible choice in low power applications due to its simple structure, achievable buck/boost voltage and easy implementation for multiple outputs. In Fig. 1, either output can be set as "master" by connecting it to the feedback sensing circuit (e.g., output 2), and the other one(s) are "slave" by adjusting the winding turn ratio (see [1] and [2]). However, differences in leakage inductances raise cross-regulation issue and increase the output errors. In industry applications, all outputs are sensed and the regulator control is based on a combination of the feedback loops. None of the outputs will be as well-regulated as the scheme where the main output has its own feedback. Therefore the total output error is nearly a constant and the error only shifts among outputs [1]. In [3-8], secondary side post regulator and low drop-out, post regulator are adopted to address the cross-regulation problem. Although the linear regulator schemes are satisfactory for tight regulation, they suffer from low efficiency because of the voltage drop across the regulator and they have limited application for low current carrying outputs. In [9-12], a magnetic amplifier is utilized to address cross regulation. However, leakage inductance and saturated inductance still affect the boundary lines and a minimum load is required to maintain the regulation.

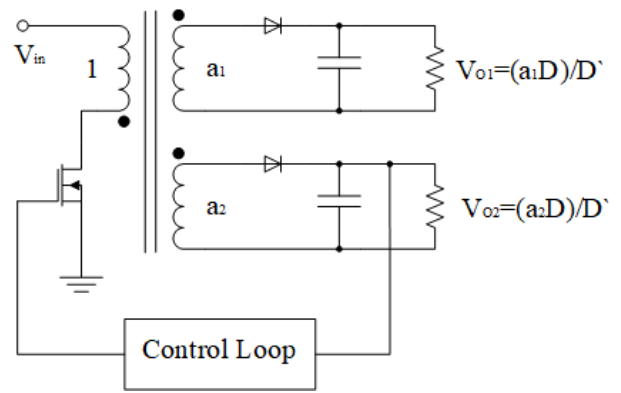

Fig. 1. Typical regulated flyback converter

This work was supported by NSF under grants ECCS-1200152.

Further efforts have been devoted in [13-18] to mitigate the cross regulation, where the need for multiple linear regulators at high load current and high frequency would deteriorate the efficiency and increase the control complexity.

This paper proposes a new single input multiple output power supply by integrating a flyback converter and several buck converters. The flyback converter works as the main regulator, and the buck converters provide series voltage compensation with the aim of tight regulation. A time multiplexing switching scheme is proposed to deliver multiple output voltage levels via a two winding transformer and to eliminate the cross regulation between output channels. This configuration reduces the size of flyback transformer and filter capacitors, and consequently improves the overall form factor. A three output channel power supply is designed and the effectiveness of the proposed configuration is validated via simulation with a MATLAB/Simscape model. Simulation results also demonstrate satisfactory transient response to load changes.

## II. Configuration Derivation

The time multiplex switching scheme applied in a boost converter in [19-21] will be adopted to regulate output voltages of flyback converter. An isolation switch is required for each channel which operates at a lower switching frequency than the main switch of the boost converter and isolates the corresponding output from other channels, during charging periods. Since the control scheme only lets one output capacitor to be charged at a time, each channel can be regulated independently, and operates at different load conditions (voltage and current) or different modes of operation (CCM or DCM). The drawback of this method appears when the number of output channels or load current is increased, as greater output capacitances are required to maintain the output voltage deviation within an acceptable range. To address this problem, a flyback-based SIMO converter is proposed in this work, which also provides galvanic isolation. Isolation is beneficial and required in many applications, e.g. EV.

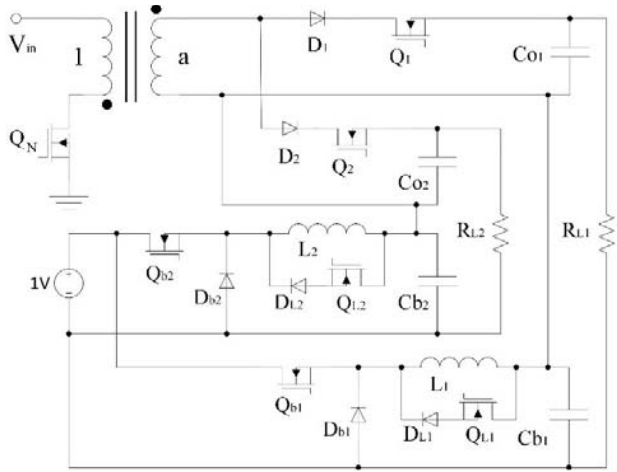

Fig. 2. Topology of proposed SIMO power supply

In addition, it allows a low power buck converter to be added in series with the flyback converter for the purpose of voltage compensation.

The proposed multiple output power supply circuit is shown in Fig. 2. As compared with conventional multiple output power supply circuit, the primary and secondary windings are sufficient to provide multiple output voltages. In fact, the form factor can be improved by removing the tertiary and quaternary windings which are essential for a typical triple output power supply. Instead, an isolation switch ($Q_1$, $Q_2$) is added to implement the time multiplex switching scheme. The low power buck converters in series with the flyback converter will be discussed later. The timing scheme is illustrated in Fig. 3. The name of switching signals are the same as that for the corresponding MOSFET, with lower case initial letter. The switching frequency ($F_s$) of the flyback and the buck converters' main switches ($Q_N$, $Q_{b1}$ and $Q_{b2}$) are equal and is set such that $F_s$ is higher than the switching frequency of the isolation switches ($F_o$). A proper $F_o$ is chosen based on a trade-off between switching loss, output capacitor size and transient response of flyback converter. In Fig. 3, the switching time period $T_o=1/F_o$ is equally divided into $N$ intervals, ($N$ is the number of output channels) and each interval assigned to one output channel.

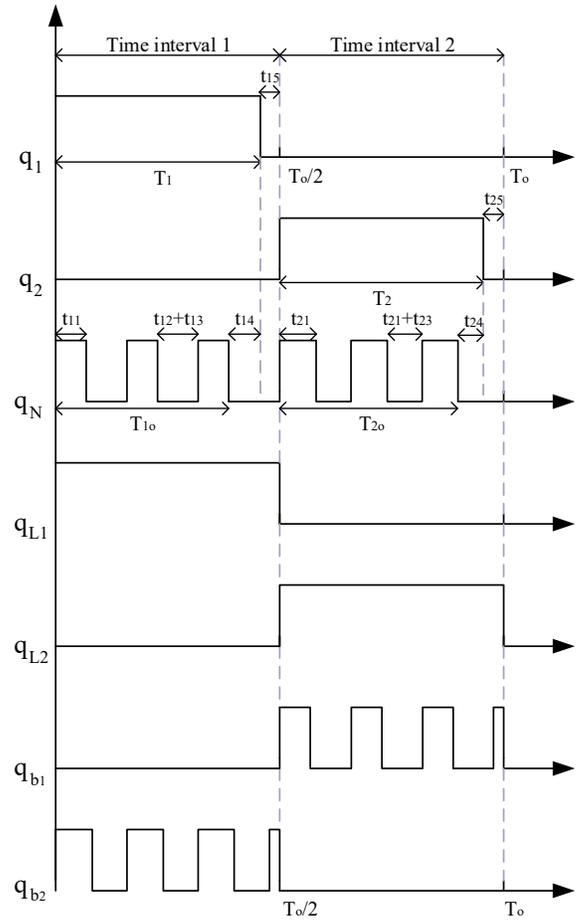

Fig. 3. Timing diagram of proposed power supply

ON-time interval of the *n*th isolation switch's switching signal ($q_n$) is set to $T_n=T_o*(\beta_n/N)$ and with a time shifting of $T_{shf}=T_o*(n-1)/N$, where *n* is the number of output channel and β*n* is always less than one for the purpose of dead time logic. It is critical for decoupling the operation of the multiple outputs, to allow the secondary charging current of the transformer reaching zero before the next output channel starts to conduct. In this work, the required time for the secondary charging current of the transformer to reach zero is called charging current reset time ($t_{14}$ and $t_{24}$). For this purpose, first the time interval that the main switch ($Q_N$) can regulate an output channel should be adjusted, which could be considered as $T_{no}=T_o*(\beta_{no}/N)$. Note that $T_n$ is always greater than $T_{no}$, since the flyback transformer stores energy when the switch $Q_N$ is ON and delivers energy to the load when this switch is OFF. By this timing scheme, the flyback converter regulates an output channel during $T_{no}$ and the stored energy is transferred to the load during $T_n$. Thus, for a double output configuration the flyback converter regulates each channel within about half of $T_o$, and then the channel will be disconnected from the flyback converter. Without series compensation the load is driven for this time interval solely by the output capacitor ($C_{on}$). This means that the output capacitor should be large enough to drive the load for the rest of the time interval which is $T_{Bn}=(N-\beta_n)*(T_o/N)$. The required capacitance can be calculated as follow:

$$C_{on} = \left(\frac{N-1}{N}\right) \cdot \frac{I_{on}}{F_o \cdot \Delta V_{on}} \quad (1)$$

Where $I_{on}$ is the *n*th channel load current and $\Delta V_{on}$ is the limit of output voltage deviation of the *n*th channel. The only parameter in (1) that limits the capacitor size is $F_o$. With increased number of outputs *(N)*, load current *($I_{on}$)* and for the applications that require tight regulation *($\Delta V_{on}$)*, normally 1% of rated output voltage, $F_o$ should be increased. The slew rate of the flyback converter would limit the highest achievable isolation switching frequency ($F_o$), otherwise the cross-regulation problem would arise. The highest achievable isolation switching frequency depends on the switching frequency ($F_s$) of the flyback converter. Higher *Fs* results in increased transformer loss and in practice it is limited within 100-500kHz. The highest achievable isolation switching frequency is reduced to 15-45kHz by many factors, including the number of output channels, the turn ON and OFF time of output channels, the time that should be saved for the secondary winding current to reach zero for each channel before starting the regulation of the next channel, and more importantly, the time that the flyback converter needs to regulate one output channel during an ON-time interval *($T_n$)*, which is at least 5-10 times of converter switching period *($T_s=1/F_s$)*. Consequently, a conventional topology without series compensation will be limited to applications with low current, high voltage and low number of output channels. The proposed topology in Fig. 2 can address the aforementioned issues, by employing the buck converters in series with output capacitors *($C_{on}$)*, which has been made possible due to the galvanic isolation of flyback converter. The input voltage of the buck converters is set to 1V, which means that it can compensate a maximal voltage drop of 1V. The voltage and current waveforms of the buck converter will be presented in simulation results section. It will be shown that the power consumption of the buck converters on average is about 1/30 of the value of the flyback converter, making on-chip implementation of the buck converters possible. When the flyback converter regulates one output channel during its charging time interval, the inductor of the corresponding buck converter is shorted out by $Q_{Ln}$ and the switching signal of the buck converter is deactivated. Once the flyback converter is connected to the next output channel, the series compensation starts to operate and stays high for $T_{Bn}$. This strategy is effective for tight regulation applications. As compared to previous configurations, smaller output capacitor can be adopted, and the power supply is regulated or compensated by closed-loop feedback controller at all time. The proposed topology eliminates cross-regulation problem, provides independent output control and achieves tight regulation for all channels.

### III. THE PRINCPLE OF OPERATIONS

In this section, the steady state of the proposed SIMO power supply with two output channels will be investigated in details. The results can be easily extended to a *n* channel topology. For simplicity, all parasitic parameters including on-resistance of switches, transformers' DC resistance and equivalent series resistance (ESR) of capacitors are disregarded. The timing scheme illustrated in Fig. 3 applies to any mode of operation (CCM/DCM). The DCM operation will be considered in the following discussion to calculate the charging current reset time, which takes greater value for DCM. Because of using a two-stage converter, two different time scales have been established, which cover operations of flyback and buck converters separately. The name of a switching signal is the same as that for the corresponding MOSFET, with lower case initial letter. Fig. 4 shows the equivalent circuit of the proposed SIMO power supply at different time intervals. In what follows, the principle of operations within *To* will be described in detail by using several equivalent circuits.

*Time interval 1 ($0 \sim T_o/N$) :*
In this time interval, when channel 1 is regulated by the flyback converter, its series buck converter is OFF to allow $C_{b1}$ be discharged by contributing to channel 1 regulation. To analyze the regulation of channel 1, some sub-intervals are established to cover all states of operation.

**Time span $t_{11}=d_{11}T_s$:**
During this time span, $Q_n$ is closed, the energy is stored in the transformer *(L*m*)*, diode $D_1$ is reverse biased, $Q_1$ is closed, $Q_{L1}$ has shorted out the inductor of the buck converter and $Q_{b1}$ and $D_{b1}$ are open. Note that diode $D_{L1}$ has been adopted to avoid unintentional current draw from MOSFET $Q_{L1}$'s body diode.

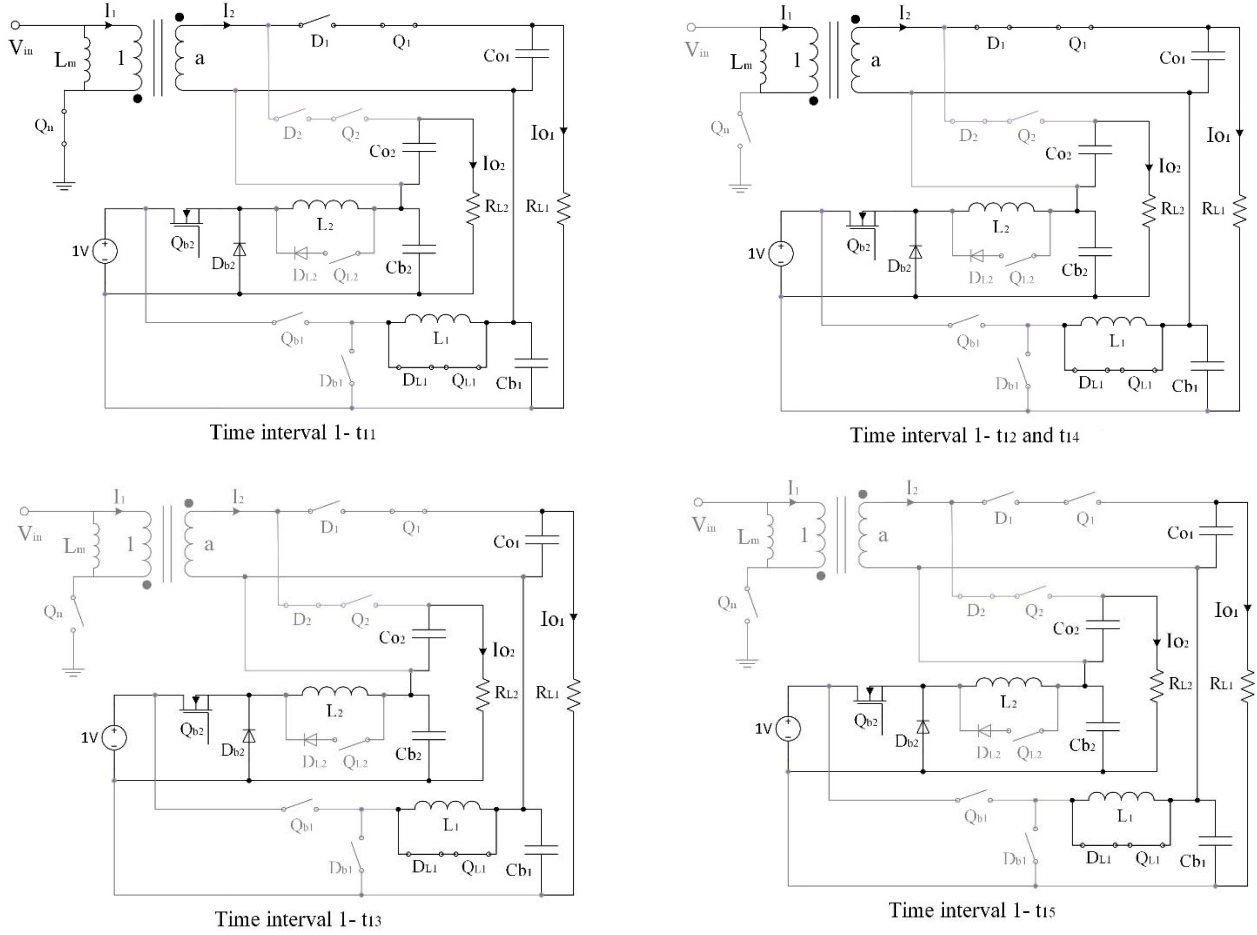

Fig. 4. Equivalent circuits of proposed SIMO power supply in time interval 1

In steady state, we have [15]:

$$I_1 = \frac{-I_2}{a} = 0 \quad (2)$$

$$\Delta I_{Lm} = I_{Lm(max)} = I_s = \frac{V_{in}}{L_m} d_{11} T_s \quad (3)$$

$$M_{VDC} = \frac{V_{o1}}{V_{in}} = \frac{d_{11}}{ad_{12}} \quad (4)$$

Where $a$ is the transformer turn ratio, $d_{11}$ is the duty cycle of the flyback converter for channel 1 and $d_{12}T_s$ is the time needed for $I_2$ to reach zero.

**Time span $t_{12}=d_{12}T_s$:**

By sending turn-OFF signal to $Q_n$, diode $D_1$ starts to conduct in this time span. $Q_1$ stays closed and allows the stored energy in $L_m$ to be transferred to the load. The equations for the current are given as below [22]:

$$I_2 = -aI_1 = aI_{Lm}$$
$$= \frac{-a^2 V_{o1}}{L_m}(t - d_{11}T_s) + \frac{aV_{in}d_{11}}{F_s L_m} \quad (5)$$

Thus:

$$I_{2(max)} = \frac{aV_{in}d_{11}}{F_s L_m} \quad (6)$$

Using (3), the dc current is computed as:

$$I_{o1} = \frac{1}{T}\int_0^T I_2 dt = \frac{1}{T}\int_{d_{11}T_s}^{(d_{11}+d_{12})T_s} I_2 dt = \frac{d_{11}a\Delta i_{Lm}}{2}$$
$$= \frac{ad_{11}d_{12}V_{in}}{2F_s L_m} = \frac{V_{o1}}{R_{L1}} \quad (7)$$

It yields

$$\frac{V_{o1}}{V_{in}} = \frac{ad_{11}d_{12}R_{L1}}{2F_s L_m} \quad (8)$$

Combining (6) and (8), we obtain

$$d_{11} = \left(\frac{V_{o1}}{V_{in}}\right)\sqrt{\frac{2F_s L_m}{R_{L1}}} \quad (9)$$

$$d_{12} = \sqrt{\frac{2F_s L_m}{a^2 R_{L1}}} \quad (10)$$

Substituting (10) in (5),

$$I_{2(max)} = aV_{o1}\sqrt{\frac{2}{F_s L_m R_{L1}}} \tag{11}$$

**Time span $t_{13}=(1-d_{11}-d_{12})T_s$:**

When $Q_n$ and $D_1$ are open, the secondary charging current $I_2$ is zero and $C_{o1}$ supplies the load. The length of this time span is $(1-d_{11}-d_{12})T_s$. At the end of this time span, one switching period of the flyback converter is completed. Completion of time interval $T_1$ occurs after repeating time spans $t_{11}$, $t_{12}$ and $t_{13}$ for $N_{itr}=T_1/T_s$ cycles.

**Time span $t_{14}$:**

Time span $t_{14}$ is actually the charging current reset time, which should be calculated based on the maximum secondary charging current $(I_{2(max)})$. The worst scenario happens when $T_{n0}$ is ended at the instant when $I_2$ reaches its maximum value $I_{2(max)}$. Note that because of the delay in microcontroller control path or other components delay, adjusting $T_{n0}$ for a specific level of $I_2$ (e.g. zero current) could be a tedious task which may vary by the time. According to (10), $t_{14}$ can be computed as,

$$t_{14} = \sqrt{\frac{2L_m}{a^2 R_{L1} F_s}} \tag{12}$$

**Time span $t_{15}$:**

Although dead time period is a short time span, it is critical to prevent cross-regulation and short circuit among output capacitors. During this time span, $Q_n$ is open to turn OFF the flyback converter, $Q_1$ and $Q_2$ are open to disconnect channels from each other and the flyback converter. When the output capacitor of channel 1 is disconnected from the converters, the series connection of $C_{o1}$ and $C_{b1}$ supply energy to the load. Here $T_n$ and $T_{n0}$ can be computed as follows:

$$T_1 = \frac{T_o}{N} - t_{15} = \frac{\beta_1 T_o}{N} \tag{13}$$

$$T_{1o} = T_1 - t_{14} = \frac{\beta_{1o} T_o}{N} \tag{14}$$

While considerable computation is required on precise timing scheme to properly control the flyback converter, the buck converter only needs to be turned ON once the flyback converter is disconnected from a channel. Although simultaneous operation of the flyback and buck converter of the same channel would not cause any problem, the optimum series compensation turn-ON timing would be after time span $t_{n5}$. The only requirement for tight regulation of channel 2, is to allow $C_{b2}$ be discharged during the turn-ON time interval of its flyback converter $(T_n)$, so that $C_{b2}$ gets maximum voltage increase during the turn-ON time interval of buck converter $(T_{Bn})$. Hence, as long as $C_{b2}$ is charged during $T_{Bn}$, the voltage decay of $C_{o2}$ would be compensated and the load voltage $(V_{o1})$ would stay tightly regulated. Step-down topology for compensator converter along with its 1v input power supply limit the voltage range of $C_{b2}$ between 0-1V. Therefore, the buck converter would never cause over-regulation, specifically when the voltage decrease of $C_{o2}$ during $T_{Bn}$ is always in effect. Channel 2 is compensated throughout $T_1$ by the series buck converter. $Q_2$ is open to disconnect the flyback converter from channel 2, while $Q_{L1}$ is open to allow the buck converter to work in normal operation by proper switching of $Q_{b2}$ and $D_{b2}$. To avoid injection of undesirable current harmonics, both converters work with the same switching frequency $F_s$.

***Time interval 2 ($T_o/N \sim 2T_o/N$) :***

In this time interval, channel 1 is compensated by its series buck converter and channel 2 is regulated by the flyback converter. The operation of the proposed SIMO power supply during this time interval is similar to that during time Interval *1* and the corresponding time spans $t_{11}$, $t_{12}$, $t_{13}$, $t_{14}$ and $t_{15}$. All presented equations and timing scheme for channel 1's flyback converter and channel 2's buck converter within time interval 1, can be adapted for channel 2's flyback converter and channel 1's buck converter, respectively. Equations for the flyback converter, which regulates channel 1 in the previous time interval, can be modified based on channel 2's circuit parameters.

IV. CIRCUIT DESIGN

In order to verify the proposed method, a triple output power supply with input voltage of 28V and rated output voltages of 15V, 18V and 30V is designed and simulated to yield 1A rated current for all channels. The voltage levels are adopted so that the flyback converter works in both step down and step up operations. The key parameters of the power supply circuit are provided in table I. The switching frequency of the flyback converter is set to 500kHz to limit the ripple voltage, which is given by [22]:

$$V_{Ripple} = \frac{r_c a D V_{in}}{F_s L_m} \tag{15}$$

Where $r_c$ is the ESR of the output capacitors. The output frequency is selected so that it provides sufficient time to regulate the output voltage. For a triple output channel power supply we have:

$$T_1 \approx \frac{F_s}{3F_o} \cdot \frac{1}{F_s} = \frac{500}{75} \cdot T_s \approx 7T_s \tag{16}$$

Therefore, the flyback converter should regulate each channel within seven switching cycles. From (1), the output capacitor of channel 3 (30V) to achieve 1% load regulation without series compensation is found to be:

$$C_{o1} = \left(\frac{2}{3}\right)\left(\frac{1}{25e3 \times 0.01 \times 30}\right) = 88\mu F \tag{17}$$

With series compensation, this capacitance is decreased by about one fourth and is selected as 25μF. The magnetizing inductance of the flyback transformer is preferred to be designed for CCM operation, since it has lower peak current and consequently lower output voltage spikes. According to [15], minimum $L_m$ can be designed for the minimum output voltage, which is 15V, as follow:

$$L_{m(min)} = \frac{a^2 V_o (1-D_{min})^2}{2F_s I_O} = \frac{15(1-0.39)^2}{2 \times 500e3 \times 1} = 5.6\mu H \tag{18}$$

The switching frequency of the buck converters is the same as that of the flyback converter to avoid EMI issue. The inductor of buck converter is designed as 2μH for CCM operation to minimize current spike and to limit the output voltage ripple. It was shown in [23] that a simple integral control can achieve practical global stability for general dc-dc converters. Hence, in the proposed control scheme for buck converters, the integration of the output voltage is used for feedback control. Simulation results in the next section illustrates that the maximum buck filter capacitor of 30μF attains 1% voltage regulation. Considering the rated buck output voltage of 1V, on-chip implementation of the series compensator could be achieved. Peak current mode control is employed to regulate the outputs of flyback converter. The peak current mode control has faster transient response and higher gain bandwidth as compared with voltage mode control, which is critical for the adopted time multiplexing switching scheme.

| Sym. | Parameters | Value |
|---|---|---|
| $F_s$ | Converters Freq. | 500kHz |
| $F_O$ | Isolation Freq. | 25kHz |
| $V_{in}$ | DC input voltage | 28V |
| $R_S$ | Sense Resistance | 0.1Ω |
| $L_m$ | Mag. inductance | 6μH |
| $L$ | Leak. inductance | 0.2μH |
| $L_{Buck}$ | Buck inductance | 2μH |
| $C_{O1}$ | Chan. 1 main Cap. | 50μF |
| $C_{b1}$ | Chan. 1 slave Cap. | 30μF |
| $C_{O2}$ | Chan. 2 main Cap. | 45μF |
| $C_{b2}$ | Chan. 2 slave Cap. | 30μF |
| $C_{O3}$ | Chan. 3 main Cap. | 25μF |
| $C_{b3}$ | Chan. 3 slave Cap. | 25μF |
| $r_C$ | Capacitor ESR | 1mΩ |
| $R_{L1}$ | Chan. 1 load | 15Ω |
| $R_{L2}$ | Chan. 2 load | 18Ω |
| $R_{L3}$ | Chan. 3 load | 30Ω |
| a | Trans. turn ratio | 1 |

Table I. Circuit Parameters

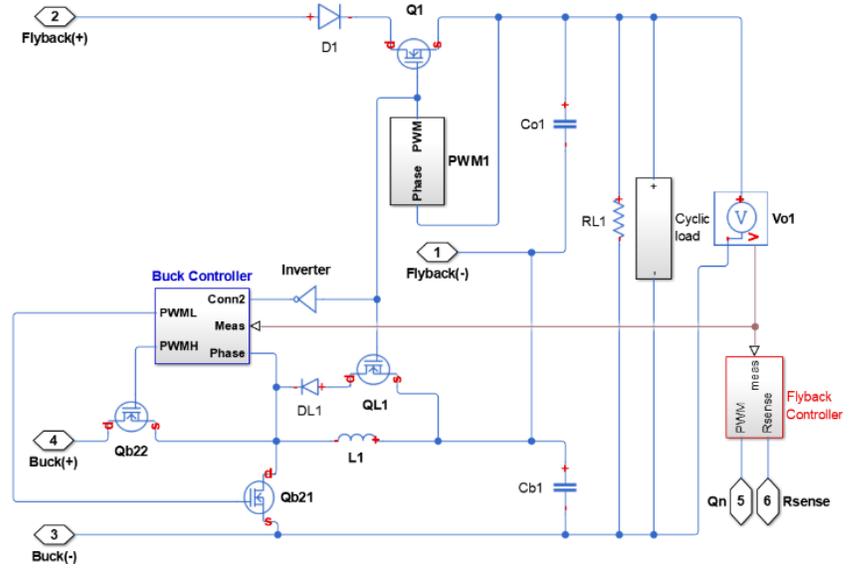

Fig. 6. Expanded view of channel 1

## V. SIMULATION RESULTS

A model for the power supply designed in the previous section is simulated with MATLAB/Simscape, which is shown in Fig. 5. Fig. 6 illustrates the expanded view of channel 1. In order to investigate the response of the power supply to the change of load, a cyclic load is incorporated in the model of Fig. 6, consisting of a controlled dc current source and a control signal. In Fig. 5, an energy regenerative snubber is used to handle the leakage inductance which causes high voltage spikes that may damage $Q_n$ when it is turned OFF. In Fig. 6, a synchronous buck converter is employed to improve the overall efficiency of the power supply circuit. PWM1 is the switching signal for the first channel isolation switch, which is also applied on $Q_{L1}$ to short out the buck inductor during the flyback operation interval. The inverted signal of PWM1 is sent to the buck converter controller to virtually turn it OFF. Fig. 7 shows the start-up response of all channels, each carrying a 1A load current. The output voltage ripple for each channel is less than 1%, achieved with output capacitances about one fourth of the conventional design. The output voltage range at steady state for the three channels are CH1=(14.91-15.13V), CH2= (17.88-18.16V) and CH3= (29.73-30.18V).

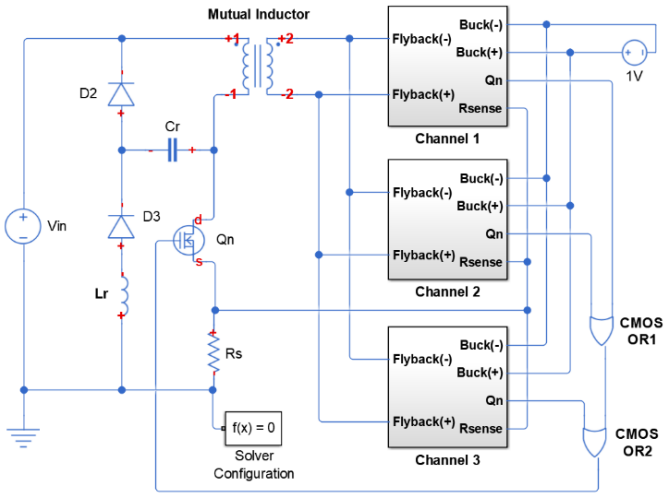

Fig. 5. Block diagram of the proposed power supply

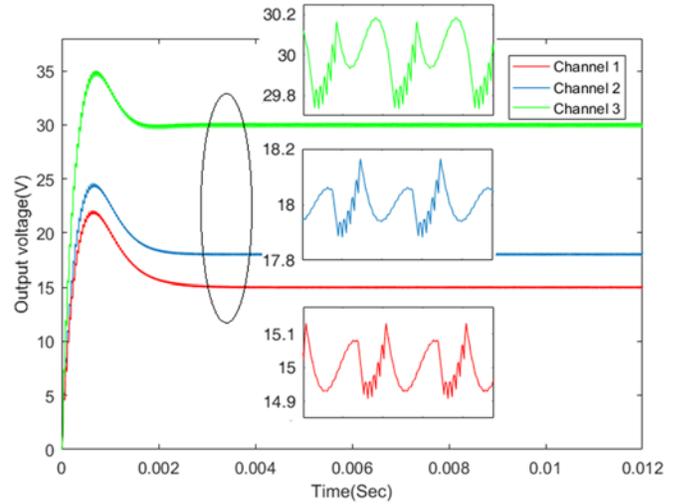

Fig. 7. Start-up response of the proposed power supply

Operations of the flyback and the buck converters are illustrated in the close-up view of Fig. 7. The thicker lines correspond to the flyback converter operation, due to the leakage inductance of the transformer. The smoother lines correspond to the operation of the buck converters. It can be seen that the third channel receives the highest compensation and the first channel receives the least. Fig. 8.a. depicts that the maximum series compensation of the first channel, inserts about 0.75V to the output capacitor, and keeps the load voltage regulated within 1%. According to (1) the channel with the smaller output voltage deviation requires higher compensation. While channel 1 is expected to have the highest compensation, it receives less inserted series voltage compared with other channels. On the other hand, the third channel, with the highest output voltage and the least voltage compensation requirement, receives the highest inserted series voltage of 0.935V. This is because the flyback output capacitor of the first channel is designed relatively larger, hence the buck converter takes the smaller portion of load regulation to stay within the desired regulation range. On the contrary, the buck converter of the third channel provides more voltage compensation, since a relatively smaller filter capacitor in the flyback converter demands more contribution from series compensation. Fig. 9 depicts the response of output voltages to the change of load, where the value of the cyclic load is 100mA for $t \in [0, 0.006]$; -100mA for $t \in (0.006, 0.012]$ and repeated periodically. The voltage responses in the figure demonstrate robust stability and desired transient response in following the reference signal. The output voltage ranges after 20% load change are CH1=(14.38-15.63V), CH2=(17.29-18.72V) and CH3=(28.93-31.05V). In other words, both the undershoots and the overshoots are kept within 4% for all channels. The regulation time for the output voltages to return within the 1% range, after a load change, is 1.75ms, 1.51ms and 1.15ms, respectively, for the 3 channels. The steady state input current of the buck converter for the first channel is depicted in Fig. 10. The maximum input current is 2.4A and its RMS value is 0.8072A.

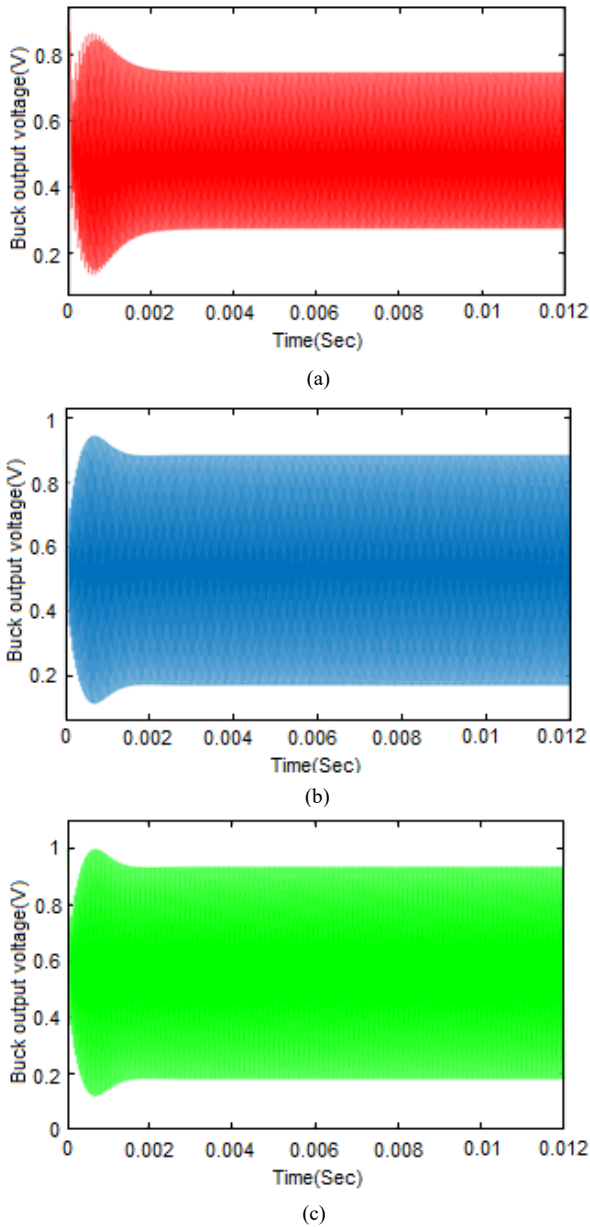

Fig. 8. Output voltage of buck converters
(a)CH1 (b)CH2 (c)CH3

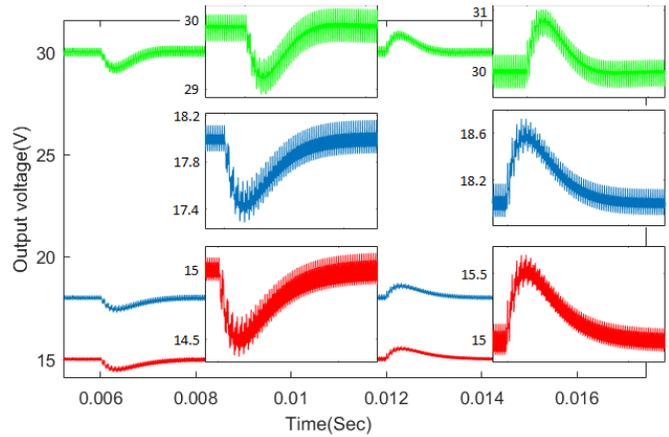

Fig. 9. Response of the power supply to 20% load change

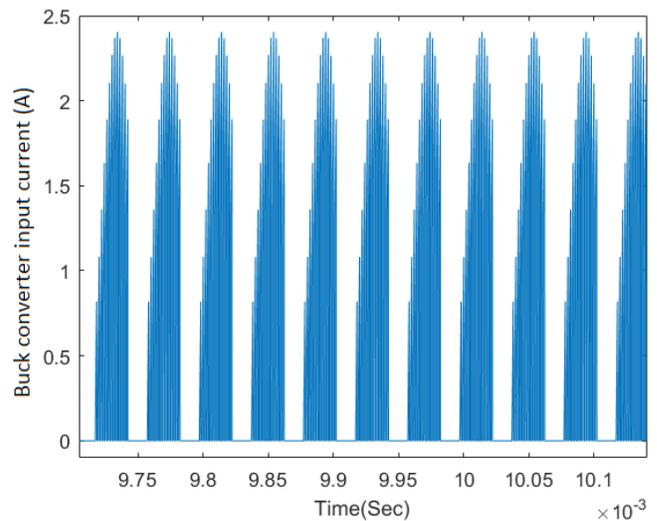

Fig. 10. Input current of channel 1's buck converter

Thus, the ratio between the first channel buck converter's input power and its rated power is:

$$\frac{P_{B1}}{P_{O1}} = \frac{V_{B(in)}I_{B1(rms)}}{V_{O1}I_{O1}} = \frac{1 \times 0.8072}{15 \times 1} \approx \frac{1}{19} \quad (19)$$

The corresponding ratio for the second and the third channels are even less;

$$\frac{P_{B2}}{P_{O2}} = \frac{V_{B(in)}I_{B2(rms)}}{V_{O2}I_{O2}} = \frac{1 \times 0.7621}{18 \times 1} \approx \frac{1}{24} \quad (20)$$

$$\frac{P_{B3}}{P_{O3}} = \frac{V_{B(in)}I_{B3(rms)}}{V_{O3}I_{O3}} = \frac{1 \times 0.7452}{30 \times 1} \approx \frac{1}{40} \quad (21)$$

The ratio between the total power consumed by the buck converters and the rated power of the power supply is:

$$\frac{P_{B(Total)}}{P_{O(Total)}} = \sum_{n=1}^{3} \frac{V_{Bn}I_{Bn}}{V_{On}I_{On}} = \frac{2.34}{63} \approx \frac{1}{27} \quad (22)$$

The total power consumed by the buck converters shows that the implementation of series compensator circuit would not change significantly the form factor of the power supply. All buck converters' components can be integrated on-chip, except for the inductors. However, using off-chip inductors for the buck converters become less of a concern in the presence of a bulky flyback transformer.

## VI. CONCLUSION

A novel multiple channel power supply is proposed in this paper by adding series compensation to a flyback converter configuration. Owing to the employed switching time scheme, one two-winding transformer suffices for a multiple output configuration. This technique achieved independent output control for each channel, eliminated cross regulation problem, reduced flyback filter capacitor size and yields tight regulation and fast transient response. The regulation error has been reduced to less than 1% with flyback capacitor size about one fourth of that in the conventional topologies. With all the desired performances, the proposed power supply circuit has smaller form factor due to low power consumption and low voltage rate of the compensator circuits.